\documentstyle[aps,prl,preprint]{revtex}
\newcommand{\beq}{\begin{equation}}
\newcommand{\eeq}{\end{equation}}

\newcommand{\beqs}{\begin{eqnarray}}
\newcommand{\eeqs}{\end{eqnarray}}
\newcommand{\beqsn}{\begin{eqnarray*}}
\newcommand{\eeqsn}{\end{eqnarray*}}

\newcommand{\nn}{\nonumber}

\begin{document}

\title{Critical value of symmetry breaking parameter in the phase transition 
 of decay rate}
\author{Hungsoo Kim$^a$, Soo-Young Lee$^b$, Sahng-Kyoon Yoo$^c$, D.K.Park$^b$,
Jae Kwan Kim$^a$}
\address{$^a$
Department of Physics, Korea Advanced Institute of Science and
Technology, \\Taejon, 305-701, Korea.\\
$^b$
 Department of Physics, Kyungnam University,
Masan, 631-701, Korea.\\
$^c$ Department of Physics, Seonam  University, Namwon, Chunbuk, 590-711, Korea.}
\date{\today}
\maketitle  
\begin{abstract}
Phase transition of decay rate from quantum tunneling to thermal activity 
regimes is investigated in (3+1)-dimensional field theories with symmetry
breaking term $f\phi$. By applying
the two independent criteria for the sharp first-order transition to the 
same model, the upper and lower bounds of critical value of the symmetry
breaking parameter are obtained. Unlike two dimensional case continuum 
states of the fluctuation operator near sphaleron solution play an important
role to determine the type of transition. 
\end{abstract}

\newpage

\newcommand{\tr}{\;{\rm tr}\;}
Recently, much attention is paid for the phase transition of the decay rate
from thermal activity to quantum tunneling regimes in (1+d)-dimensional
field theories when $d=1$[1], $d=2$[2], and $d=3$[3]. Especially, 
$d=3$ case, which is a concern of the present letter, has an abundant applications
in cosmology[4] and particle physics[5, 6]. In this problem
potential is usually choosed, for simplicity, as an asymmetric double well potential
\beq
V_1[\phi] = \frac{1}{2} (\phi^2 - 1)^2 - f \phi,
\eeq
where $f$ is symmetry breaking parameter whose range is 
$0 < f < 4/3\sqrt{3}$. In fact, Eq.(1) is a tree level potential of the 
temperature-dependent corrections[7,8]. The analysis of the phase transition
for the temperature-dependent effective potential by the method used in this
letter is expected to be very complicated and will be discussed elsewhere.
 After taking an appropriate scaling transformation
it is easy to show that the system with potential (1) exhibits a same 
physics with a system whose potential is[9]
\beq
V_2[\phi] = \frac{1}{4} \phi^4 - \phi^3 + \frac{1}{2} \delta \phi^2,
\eeq
if $\delta$ has a relation with $f$ as follows:
\beq
f = \frac{2(2 - \delta)}{(3 - \delta)^{\frac{3}{2}}}.
\eeq
From the range of $f$ one can see that the parameter $\delta$ can vary between
$0$ and $2$. In the limit $\delta \rightarrow 2$(or $f \rightarrow 0$),
we can use the usual thin wall approximation method[10]. In the opposite
limit $\delta \rightarrow 0$(or $f \rightarrow 4 / 3\sqrt{3}$), the height 
of the barrier between the two vacua vanishes. Since the radius of bubble
is smaller than the wall thickness, this limit is frequently called thick wall
limit[4].

In Ref.[3] Ferrera has shown that the type of the phase transition of 
decay rate is sharp first-order when $f = 0.25$ and $f = 0.55$, and smooth
second-order when $f = 0.75$ by solving the equation of motion
numerically with an aid of the multigrid method.
This means that there exists a critical value  
$f_{\ast}$(or $\delta_{\ast}$), which distinguishes the type of 
transition. The multigrid method, however, is not useful to determine 
$f_{\ast}$(or $\delta_{\ast}$) because it is impossible to treat temperature
continuously due to the fixed grids. Furthermore, it may be highly difficult
to solve the equation of motion numerically in realistic models.
It is, therefore, important to determine $f_{\ast}$ without solving equation
of motion. It may be worthwhile noting that the terminology ``phase transition''
used in the present letter does not mean the generic transition of the 
physical system's phase but the transition between the two different instanton
regimes. 

In this letter we will evaluate the lower and upper bounds of 
$f_{\ast}$(or $\delta_{\ast}$) by using two independent criteria
for the sharp transition.

Let us start with Euclidean action
\beq
S_E[\phi] = \int d^4x 
\left[ \frac{1}{2} 
      \left( \frac{\partial \phi}{\partial \tau} \right)^2
      + (\vec{\bigtriangledown} \phi)^2 + V_2[\phi]  \right],
\eeq
which yields an equation of motion
\beqsn
\frac{\partial^2 \phi}{\partial \tau^2} + 
\vec{\bigtriangledown}^2 \phi = V_2^{\prime}[\phi].
\eeqsn
Since zero temperature bounce solution $\phi_B$ has $O(4)$ symmetry[11],
the equation of motion for $\phi_B$ is reduced to 
\beq
\frac{d^2 \phi_B}{d R^2} + \frac{3}{R} \frac{d \phi_B}{d R}
= V_2^{\prime}[\phi_B],
\eeq
where $R = \sqrt{\tau^2 + \vec{x}^2}$. Thus, the classical action for $O(4)$ symmetric
solution is 
\beq
S_4 = 2 \pi^2 \int_0^{\infty} dR R^3
\left[ \frac{1}{2} \left( \frac{d \phi_B}{dR} \right)^2 + V_2[\phi_B] \right].
\eeq

At high temperature thermal activity is governed by static sphaleron
solution[12] which has a $O(3)$ symmetry. Hence, the equation of motion for the 
sphaleron solution $\phi_{sph}$ is 
\beq
\frac{d^2 \phi_{sph}}{dr^2} + \frac{2}{r} \frac{d \phi_{sph}}{dr}
= V_2^{\prime}[\phi_{sph}],
\eeq
where $r = \sqrt{\vec{x}^2}$, and, thus, the classical action for this 
configuration is 
\beq
S_{sph} = \tau_{sph} S_3.
\eeq
Here, $\tau_{sph}$ is the period of small fluctuation around $\phi_{sph}$ and 
\beq
S_3 = 4 \pi \int_0^{\infty} r^2
\left[ 
\frac{1}{2} \left( \frac{d \phi_{sph}}{dr} \right)^2 + V_2[\phi_{sph}]
                                                                 \right].
\eeq
At a finite temperature $T$ the solution with period $1/T$ has an information
on the decay rate[4, 13], which is a result of the
saddle point approximation.

 The possible type of transition of decay rate
from thermal activity to quantum tunneling regimes is thoroughly discussed
in Ref.[14] within quantum mechanical models. By using a relation
\beq
\frac{d S_E}{d \tau} = E,
\eeq
where $\tau$ and $E$ are the period of solution
 and energy, respectively, Ref.[14] has shown
that the Euclidean action for the finite temperature solution meets 
$S_3 / T$-curve smoothly at $T_{sph} = 1 / \tau_{sph}$. 
The general features
of action-vs-temperature for the first- and second-order transitions are 
shown at Fig.1.
Although it is not general criterion for the sharp transition,
 one can see from Fig.1(b) that $S_{sph}$ is 
larger than $S_4$ for the comparatively strong first-order transition.

In spite of quantum mechanical ground of Ref.[14] its generalization to 
field theories is straightforward since Eq.(10) holds even in field theories
as a saddle point equation[15]. So, we can use a condition 
$S_{sph} > S_4$ for a strong first-order transition.
Since $S_3$ and $S_4$ have been obtained in Ref.[9] numerically
\beqs
S_3&=&\frac{64 \pi}{81} (2 - \delta)^{-2} \sqrt{\delta}
      (\beta_1 \delta + \beta_2 \delta^2 + \beta_3 \delta^3), \nn \\ 
S_4&=& \frac{4 \pi^2}{3} (2 - \delta)^{-3} 
      (\alpha_1 \delta + \alpha_2 \delta^2 + \alpha_3 \delta^3),
\eeqs
where $\beta_1=8.2938$, $\beta_2=-5.5330$, $\beta_3=0.8180$, 
$\alpha_1=13.832$, $\alpha_2=-10.819$, and $\alpha_3=2.0765$, the remaining
one we have to calculate is $\tau_{sph}$.

To obtain $\tau_{sph}$ let us consider the small thermal fluctuation around
the sphaleron:
\beq
\phi(\tau, \vec{r}) = \phi_{sph}(r) + \eta(\tau, \vec{r}).
\eeq
Inserting it to the field equation, one can show directly that 
$\eta(\tau, \vec{r})$ obeys
\beq
\hat{l} \eta = \hat{h} \eta + \hat{G}_2[\eta] + \hat{G}_3[\eta],
\eeq
where
\beqs
\hat{l}&=&\frac{\partial^2}{\partial \tau^2}, \nn \\  
\hat{h}&=&-\vec{\bigtriangledown}^2 + (3 \phi_{sph}^2 - 6 \phi_{sph} + \delta), 
\nn\\ 
\hat{G}_2[\eta]&=&3 (\phi_{sph} - 1) \eta^2,  \\
\hat{G}_3[\eta]&=& \eta^3. \nn
\eeqs
After neglecting $\hat{G}_2[\eta]$ and $\hat{G}_3[\eta]$ which are higher
order terms of $\eta$, one can show easily
\beq
\tau_{sph} = \frac{2 \pi}{\sqrt{\mid h_0 \mid}},
\eeq
where $h_0$ is negative eigenvalue of $\hat{h}$ operator.
Since eigenfunction for the negative eigenvalue is angle-independent, we have 
to solve the spectra in the radial equation
\beq
\left[- \frac{d^2}{dr^2} - \frac{2}{r} \frac{d}{dr} + U(r) \right]
u_n(r) = h_n u_n(r),
\eeq
where $U(r) = 3 (\phi_{sph}^2 - 2 \phi_{sph}) + \delta$.
The eigenvalues for the negative mode and first positive mode are obtained 
numerically(See Fig.2).
Fig.3 shows the numerical result of $S_{sph}/S_4$ when Eq.(11) is used.
In this case we get a condition $\delta > \delta_1^A = 1.45$ for
the strong first-order transition. Since, however, the author of Ref.[9] obtained
$S_3$ and $S_4$ by fitting formula with only three parameters, Eq.(11)
can be rough estimation. So, we calculated again $S_3$ and $S_4$
numerically by using the simple shooting method, which results in
$\delta > \delta_1 =1.42$, which is lying within his fitting error,
 as shown in Fig.3.

The opposite bound is obtained by using a criterion for the first order 
transition developed at Ref.[16]. In fact, authors of Ref.[16] obtained
the condition for the sharp transition analytically which can be summarized
as follows:
\beq
<u_0 \mid f[u_0]>\,\,\,\, < \,\,\,0,
\eeq
where 
\beqs
f[u_0] =&-& \frac{1}{2} \frac{\delta \hat{G}_2}{\delta \eta}\bigg|_{\eta = u_0}
 \left[\hat{h}^{-1} + \frac{1}{2} (\hat{h} + 4 \omega_{sph}^2)^{-1} \right]
\hat{G}_2[u_0]  \\  \nn
&+& \frac{3}{4} \hat{G}_3[u_0],
\eeqs
$<a \mid b>$ is usual inner product, and  $u_0$ is negative mode of 
$\hat{h}$.
Noting that the angle dependent eigenfunctions do not contribute to 
$<u_0 \mid f[u_0]>$ and using explicit forms of $\hat{G}_2$ and 
$\hat{G}_3$ given at Eq.(14), one can reduce the condition (17) in our case to
\beq
I_1 + I_2 < 0,
\eeq
where
\beqs
I_1&=& \frac{3}{4} <u_0 \mid \hat{G}_3[u_0]> - 
\left( \frac{1}{h_0} + \frac{1}{2} \frac{1}{h_0 + 4 \omega_{sph}^2} \right)
\mid <u_0 \mid \hat{G}_2[u_0]> \mid^2,  \nn \\
I_2&=& - J_1 - J_2.  
\eeqs 
Here, 
\beqs
J_1&=& \sum_{n \geq 1}
\left( \frac{1}{h_n} + \frac{1}{2} \frac{1}{h_n + 4 \omega_{sph}^2} \right)
\mid <u_n \mid \hat{G}_2[u_0]> \mid^2,  \\
J_2&=&\int dk
\left( \frac{1}{h_k} + \frac{1}{2} \frac{1}{h_k + 4 \omega_{sph}^2} \right)
\mid <u_k \mid \hat{G}_2[u_0]> \mid^2. \nn
\eeqs
$u_0$, $u_n (n = 1, 2, \cdots)$, and $u_k$ are negative mode, discrete
positive modes, and continuum states in the radial equation (16), 
respectively, and $h_0$, $h_n$, and $h_k$ are corresponding eigenvalues.
Since $I_1 > 0$ and $I_2 < 0$, the competition of $I_1$ and $I_2$ determines  
the type of transition.
Numerical calculation shows that there exists only one positive discrete mode
in the radial eigenvalue equation (16). Hence, $J_1$ is simply
\beq
J_1 = 
\left( \frac{1}{h_1} + \frac{1}{2} \frac{1}{h_1 + 4 \omega_{sph}^2} \right)
\mid <u_1 \mid \hat{G}_2[u_0]> \mid^2.
\eeq
Although the criterion for the sharp transition (19) is analytic expression,
it is highly difficult to treat the continuum states numerically. In order to 
escape this difficulty we change $J_2$ to its upper bound $J_2^M$
\beqs
J_2 
&\rightarrow& J_2^M \nn \\  
&=& \left( \frac{1}{\delta} + \frac{1}{2} \frac{1}{\delta + 4 \omega_{sph}^2}
                                                                  \right)
     \int dk \mid <u_k \mid \hat{G}_2 [u_0]> \mid^2 \nn \\
&=& \left( \frac{1}{\delta} + \frac{1}{2} \frac{1}{\delta + 4 \omega_{sph}^2}
                                                                  \right)
       \\
&& \times
\left[ <\hat{G}_2[u_0] \mid \hat{G}_2[u_0]> 
      - \mid <u_0 \mid \hat{G}_2[u_0]> \mid^2 
    - \mid <u_1 \mid \hat{G}_2[u_0]> \mid^2  \right].\nn
\eeqs
In deriving the upper bound  $J_2^M$, we used the fact that the 
continuum eigenvalue is $\delta + k^2$.
Since $J_2^M > J_2$, $I_1 + I_2^M > 0$ is a sufficient condition
 for the second-order
transition, where $I_2^M = - J_1 - J_2^M$.
Results of numerical calculation for $I_1 -J_1$ and $I_1-J_1-J_2^M$ are given at Fig.4.
Since  $I_1 -J_1$ is the contributions of only negative and discrete positive
modes to $<u_0\mid f[u_0]>$, the fact that $I_1-J_1 > 0$ in the full range
of $\delta$ implies the contribution of the continuum states are very important
for the occurrence of the sharp first-order transition. This is new feature
which does not arise at (1+1)-dimensional case, in which the contribution
of the continuum states is negligible[16]. Hence, we get $\delta < \delta_2=1.0245$
for the second-order transition. 
Combining Fig.3 and Fig.4 one can conclude $1.0245 < \delta_{\ast} < 1.42$
or in terms of $f$ $0.584 < f_{\ast} < 0.703$.
Fig.5 shows bubble nucleation at $\delta =1.18$ which is calculated by the 
multigrid method, which shows the existence of the wiggly solution in the
intermediate range of temperature.

In this letter, we derived the lower and upper bounds of $f_{\ast}$ or
 $\delta_{\ast}$
by applying the two independent criteria $S_{sph} > S_4$ and $I_1 + I_2 < 0$
to the same system simultaneously.
Although one can obtain more accurate value of $f_{\ast}$ by solving equation
of motion numerically in the simple model, this might be impossible for more
complicated models like electroweak theory and real cosmological models whose
potential are generally dependent on temperature explicitly.
In these cases our method presented in this letter would be useful to determine
the type of transition and the critical values of some parameters involved
in the models.

\newpage 
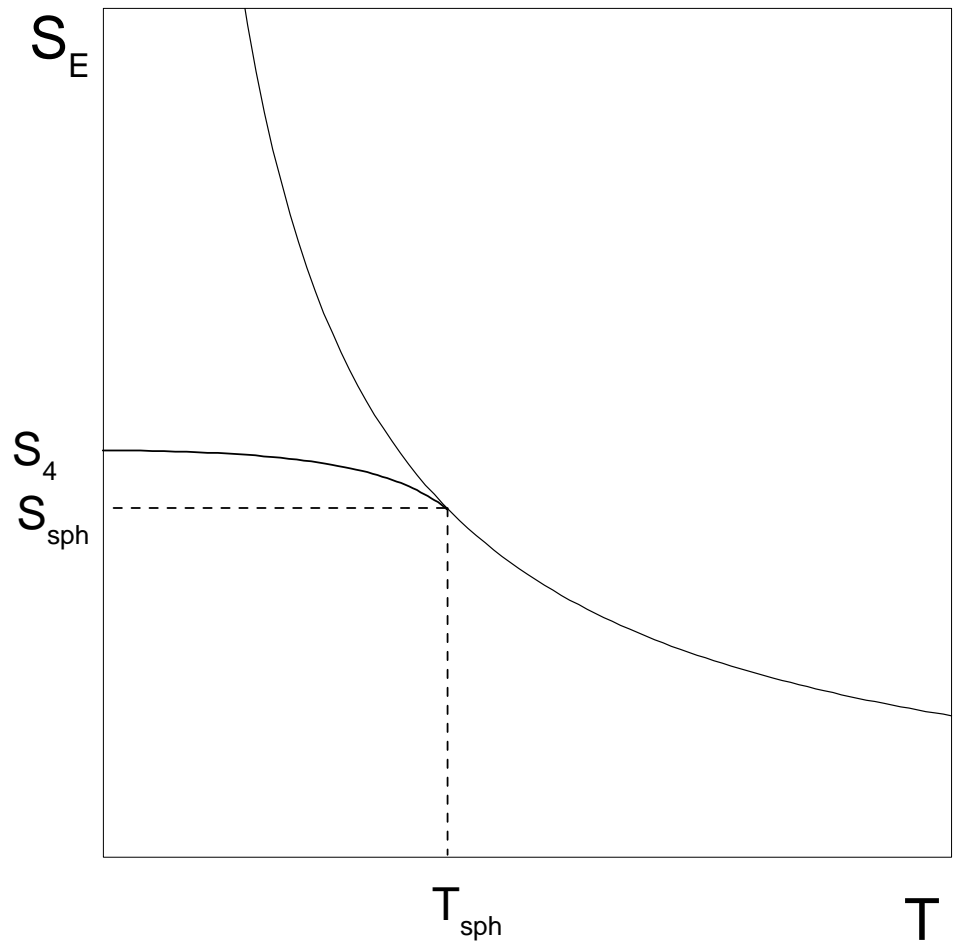
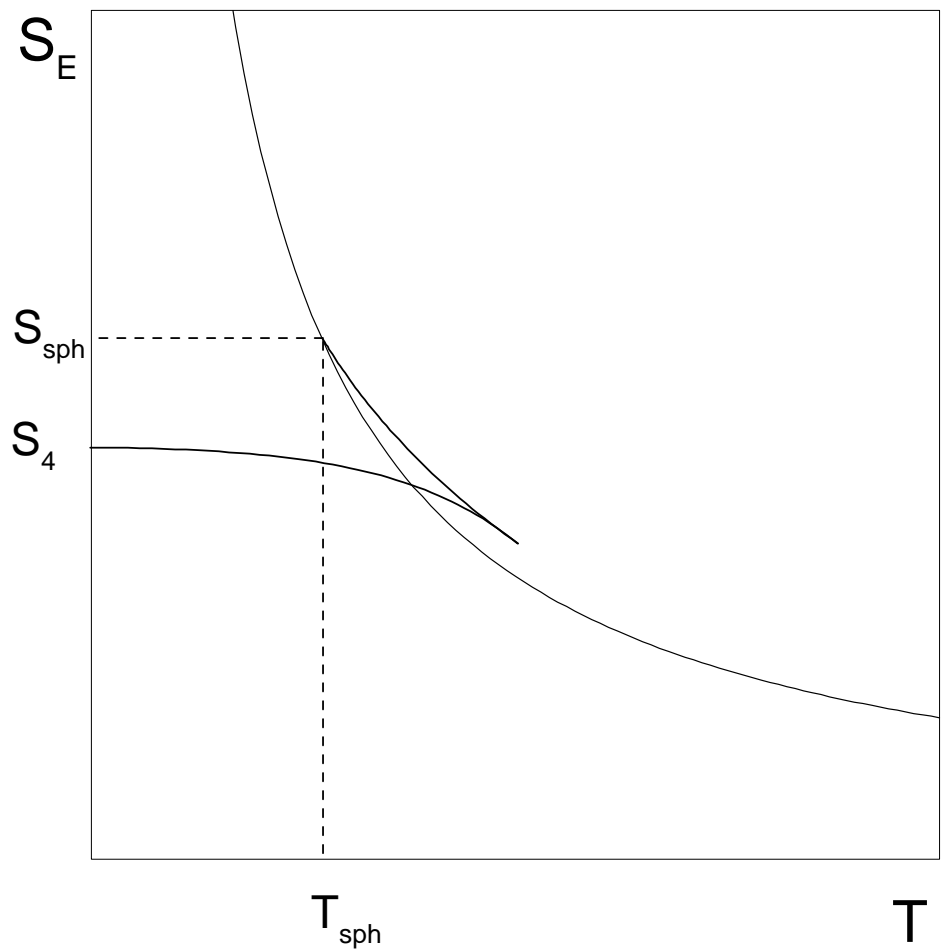
\begin{figure}
\caption{(a) Second-order transition from the thermal to the quantum regimes.
(b) First-order transition from the thermal to the quantum regimes. }
\end{figure}
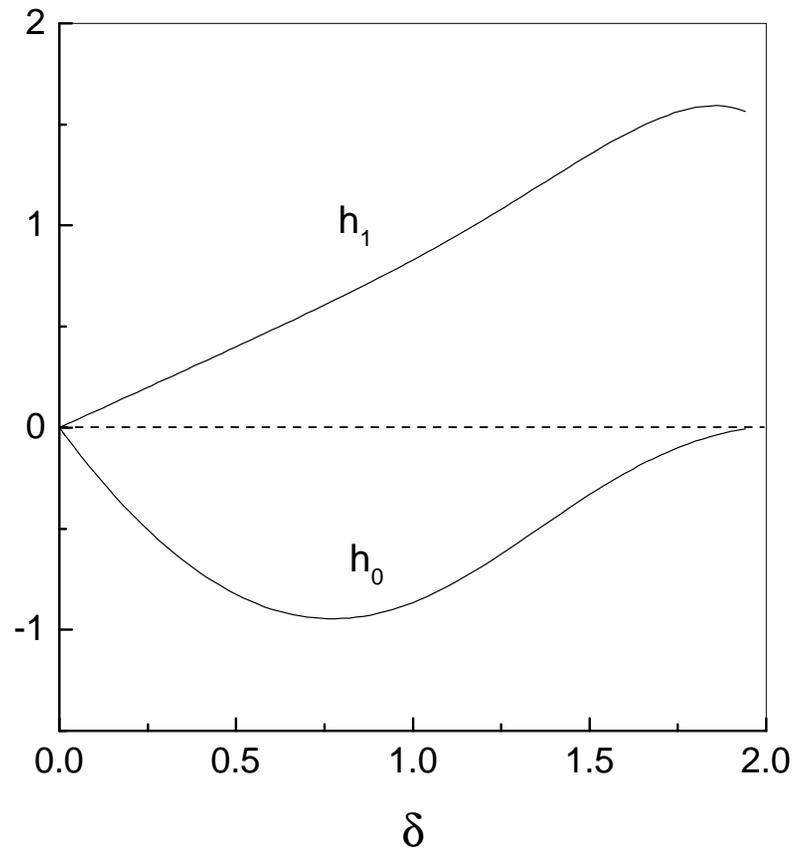
\begin{figure}
\caption{Negative($h_0$) and discrete positive($h_1$) eigenvalues of $\hat{h}$
operator.}
\end{figure}
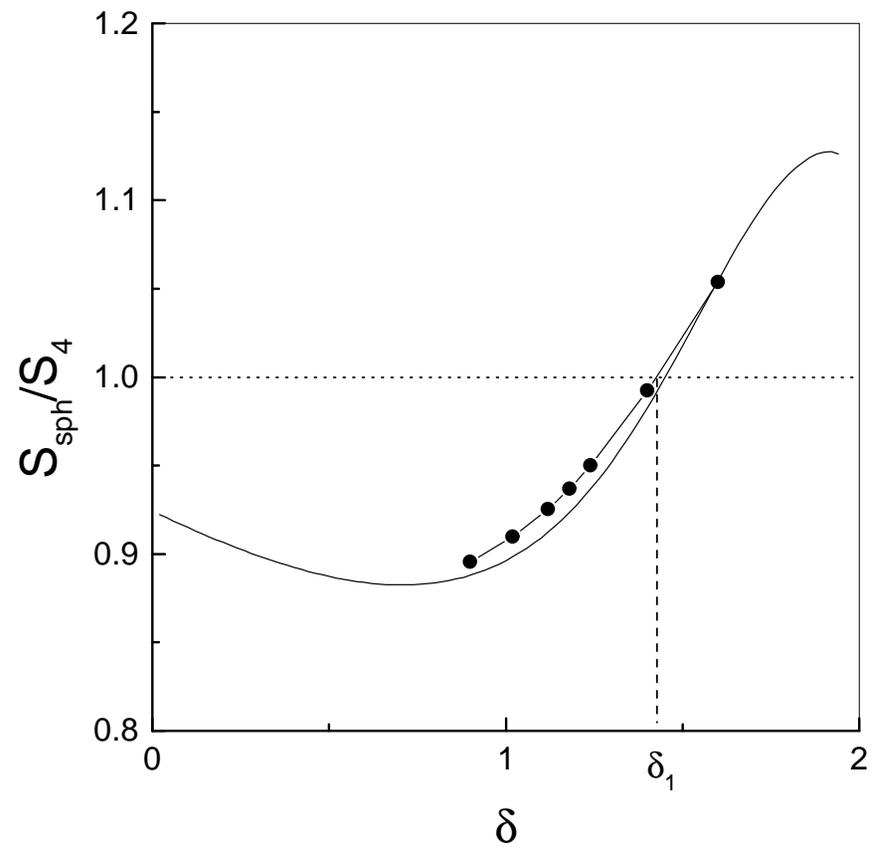
\begin{figure}
\caption{Plot of $S_{sph}/S_4$ with respect to $\delta$.
The solid line is calculated by using Eq.(11) and the data points
are obtained by shooting method. In the range of $\delta$ corresponding
to  $S_{sph}/S_4 > 1$ the transition of the decay rate becomes comparatively
strong first-order.}
\end{figure}
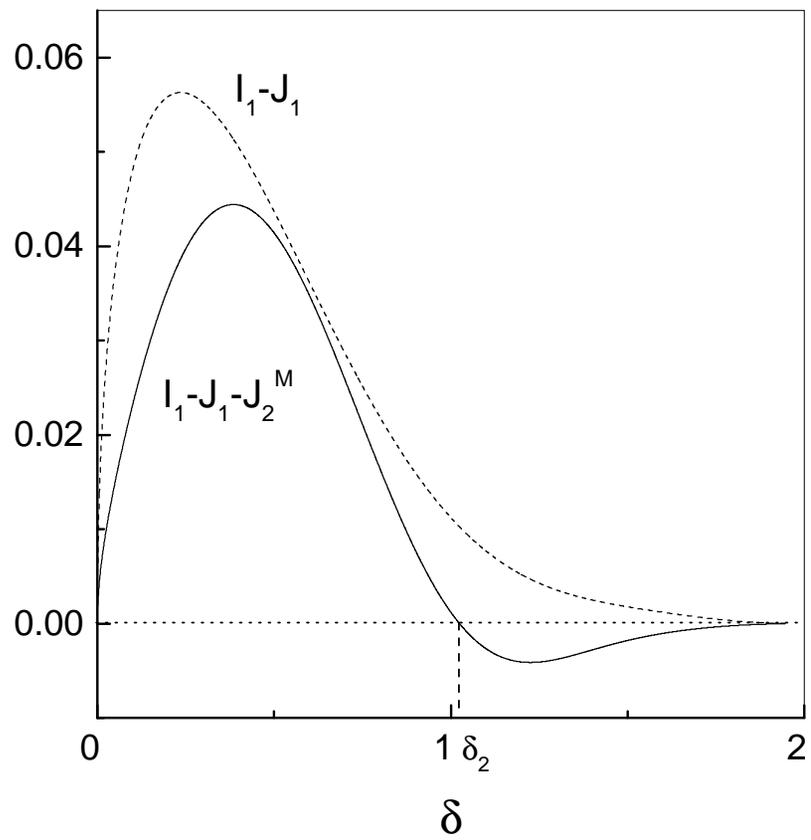
\begin{figure}
\caption{Plots of $I_1-J_1$ and $I_1-J_1-J_2^M$ with respect to $\delta$. 
The fact that $I_1-J_1 > 0$ in the full range of $\delta$ means that
the continuum states of operator $\hat{h}$ play an important role
for the determination of the type of transition. From the plot $I_1-J_1-J_2^M$
we can conjecture that the transition becomes smooth second-order when 
$\delta < \delta_2$.}
\end{figure}
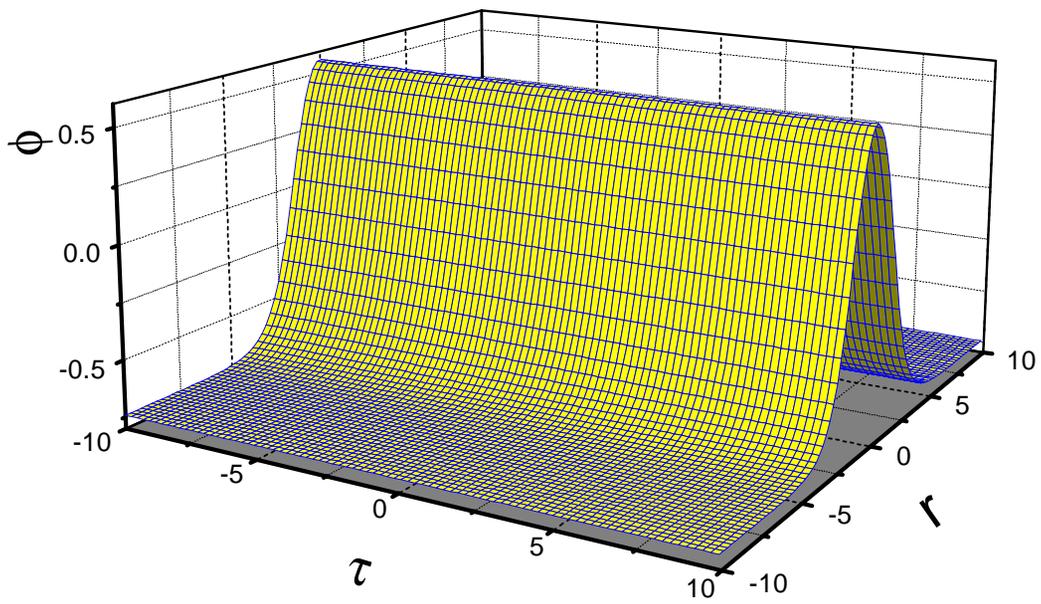
\begin{figure}
\caption{Bubble nucleation at $\delta =1.18$.
(a) Periodic bubble solution at $T=0.763 T_{sph}$.
(b) Periodic wiggly solution at $T=0.994 T_{sph}$.
(c)$O(3)$ symmetric sphaleron solution at high temperatures.}
\end{figure}

\end{document}